\documentclass [12pt,preprint]{aastex}

\usepackage{epsfig}

\begin{document}

\newcommand{\gsim}{\hbox{\rlap{$^>$}$_\sim$}}

\title{Is the afterglow of Gamma Ray Burst  021004 unusual?}

\author{Shlomo Dado\altaffilmark{1}, Arnon Dar\altaffilmark{1,2} and
A. De R\'ujula\altaffilmark{2}}

\altaffiltext{1}{Physics Department and Space Research Institute,
Technion, Haifa 32000, Israel}
\altaffiltext{2}{Theory Division, CERN, CH-1211 Geneva 23, Switzerland}


\begin{abstract}

The bumpy light-curve of the bright optical afterglow (AG) of gamma ray
burst (GRB) 021004, its spectral evolution and its radio emission have
been claimed to be unusual.  In the Cannonball Model of GRBs that is not
the case.  The very early AG's shape is, as for GRB 990123, a direct
tracer of the expected circumburst density profile. The unprecedented
precision of the data allows for the ``resolution'' of two cannonballs
(CBs) in the AG. These two CBs correspond to the two pulses in the GRB and
to the two wide humps in the AG light curve. The smaller wiggles in the AG
are, as for GRBs 000301c and 970508, to be expected:  they trace moderate
deviations from a constant density interstellar medium.  The observed
evolution of the optical spectrum is that predicted in the CB model. The
X-ray and radio emissions of GRB 021004 are also normal.

\end{abstract}


\section*{Introduction}

Instruments aboard HETE II detected GRB 021004 at 4.50432 UT October
2002 (Shirasaki et al. 2002). Its bright fading optical afterglow
was discovered by Fox et al. (2002) nine minutes after the
burst and was soon followed worldwide with many telescopes, quite
continuously during the first few days. The unprecedented density
and precision of the data are reported in Fig. \ref{fone}.  The
redshift of the GRB's host galaxy, $\rm z=2.33$, was first correctly
determined by Chornock and Filippenko (2002) and confirmed by
Salamanca et al.  (2002), Mirabal et al.  (2002a), Savaglio et al.
(2002), Castro-Tirado et al. (2002) and Matheson et al. (2002).
The AG's brightness allowed precise measurements of its temporal
decline, polarization, spectrum, and spectral evolution.  The
optical light-curves deviate from the smoothly steepening decline
observed in most well-studied optical AGs. There are also
``unexpected'' features in the spectrum (e.g., Salamanca et al.
2002; Mirabal et al. 2002b)  and its evolution (e.g., Matheson et
al. 2002).  The spectrum of the radio to submillimeter wavelengths
was also claimed to be ``unusual'' (Berger et al. 2002a).

In this letter we show that --in the CB model-- the only uncommon
feature of GRB 021004 is the precision of the measurements.  Its
most ``surprising''  properties were anticipated, in the
sense that there are --again, in the CB model-- precedents for all of them
in past GRBs.

\section{The Cannonball Model of GRBs}

There is mounting evidence that long duration GRBs are produced in
the explosions of supernovae akin to SN1998bw (Galama et al. 1998),
by the ejection of ordinary baryonic matter --essentially ionized
Hydrogen-- in the form of plasmoids or ``cannonballs'' (CBs), with
very highly relativistic Lorentz factors ($\gamma\sim 10^3$) (Dar
and De R\'ujula 2000, 2001, Dado et al. 2002a,b,c,d) but otherwise
similar to the ones observed in quasars (Marscher et al. 2002) and
microquasars (e.g., Mirabel and Rodriguez 1994;  Belloni et al.
1997; Rodriguez and Mirabel 1999; Mirabel and Rodriguez 1999;
Corbel et al. 2002, and references therein).

In the CB model  (Dar and De R\'ujula 2000, 2001, reviewed in De R\'ujula
2002a,b) long duration GRBs and their AGs are produced in core
collapse SNe by jets of highly relativistic CBs that pierce through
the SN shell and the SN progenitor's ``wind'' ejecta. A CB is
emitted, as observed in $\mu$-quasars, when part of an accretion
disk falls abruptly onto the newly-born compact central object.
Crossing the circumburst shells with a large Lorentz factor $\gamma$,
the surface of a CB is collisionally heated to keV temperatures
and the radiation it emits when it reaches the transparent outskirts
of the shells ---boosted and collimated by the CB's motion--- is
a single $\gamma$-ray pulse in a GRB. The cadence of pulses reflects
the chaotic accretion and is not predictable, but the individual-pulse
temporal and spectral properties are. A long list of general
properties (Dar and De R\'ujula 2001) of GRB pulses is reproduced
in the CB-model, in which, unlike in the standard ``fireball''
models, the GRBs' $\gamma$'s have a thermal-bremsstrahlung
---as opposed to synchrotron--- origin\footnote{In this
(Ghirlanda et al. 2002), as in the relevance of the
viewing angle, or the explanation of AG features by density inhomogeneities,
the ``standard'' model is phagocytizing the CB-model's views.}.

A CB exiting the circumburst shells soon becomes transparent to
its own enclosed radiation. At that point it is still expanding
and cooling adiabatically and by bremsstrahlung. The hard bremsstrahlung
spectrum dominates the very early X-ray AG (for a few tens of
seconds) with a fluence of predictable magnitude decreasing with
time as $\rm t^{-5}$.  All X-ray AGs are compatible in magnitude
and shape with this prediction (Dado et al. 2002a).

After the first tens of seconds, a CB's emissivity is dominated by
synchrotron emission from the electrons that penetrate in it
from the interstellar medium (ISM).
Integrated over frequency, this synchrotron emissivity is proportional
to the energy-deposition rate of the ISM electrons in the
CB\footnote{The kinetic energy of a CB is mainly lost to the ISM
protons it scatters; only a fraction $\rm\leq m_e/m_p$ is re-emitted
by the incorporated electrons, as the AG.}. These electrons are
Fermi accelerated
in the CB's tangled magnetic maze to a broken power-law energy
distribution with a ``bend'' energy
 equal to their incident energy in the CBs' rest frame. In that
frame, the electrons' synchrotron emission (prior to attenuation
corrections) has an approximate spectral energy density  (Dado et
al.  2002b):
\begin{equation}
\rm  F_{_{CB}}[\nu,t] \simeq \rm f_0\,
{ (p-2)\, \gamma^2\, n_p\, m_e\, c^3 \;  [\nu/\nu_b]^{-1/2}
\over (p-1)\,\nu_b\, \sqrt{1+[\nu/ \nu_b]^{(p-1)}}}\, ,
\label{sync}
\end{equation}
where $\rm f_0$ is a normalization constant,
$\rm p\approx 2.2$ is the spectral index of the Fermi
accelerated electrons prior to the inclusion of radiation losses,
$\rm n_p$ is the ISM baryon density, $\rm
\gamma(t)=1/\sqrt{1-\beta^2}$ (with $\rm \beta=v/c$) is the Lorentz
factor of the CBs, and $\rm \nu_b $
is the ``injection bend'' frequency in the CB's rest frame\footnote{This
bend frequency is  not  the conventional
``cooling break''. It reflects an {\it injection bend} in
the  electron spectrum  at the energy $\rm E_b
= \rm\gamma(t)\ m_e\, c^2$ with which the ISM electrons enter the
CB at a particular time in its decelerated motion (Dado et al.
2002b).},
\begin{equation}
\rm \nu_b \simeq \rm 1.9\times 10^{3}\,[\gamma(t)]^3\;
\left[{n_p\over 10^{-3}cm^{-3}}\right]^{1/2} \;\; Hz.
\label{nubend}
\end{equation}

The radiation emitted by a CB is Doppler-shifted and
forward-collimated by its highly relativistic motion, and redshifted
by the cosmological expansion.  A distant observer sees a spectral
energy flux:
\begin{equation}
\rm F_{obs}[\nu,t]\simeq
\rm { (1+z)\,\delta(t)^3\, R^2\, A(\nu,t)\,\over D_L^2}\,
F_{_{CB}}[\nu',t']\, ,
\label{Fnuobser}
\end{equation}
where $\rm R$ is the radius of the CB (which in the CB model tends
to a calculable constant value ${\rm
R_{max}}$ of ${\cal{O}}(10^{14})$ cm, in minutes of observer's time),
$\rm A(\nu,t)$ is the total extinction along the line of sight to
the CB, $\rm D_L(z)$ is the luminosity distance (we
adopt $\rm H_0=65$ km/(s Mpc), ${\rm
\Omega_M}=0.3$ and ${\rm \Omega_\Lambda}=0.7$),
$\rm \nu'=(1+z)\,\nu/\delta(t)$,  $\rm t'=\delta(t)\,t/(1+z)$,
and $\rm\delta(t)$
is the Doppler factor of the light emitted by the CB.
In the domain of interest for GRBs, $\rm \gamma^2\gg 1$ and
$\theta^2\ll 1 $, with $\theta$  the angle between the CB's
direction of motion and the line of sight to the observer,
\begin{equation}
\rm \delta(t)
\simeq{2\, \gamma(t)\over 1+\theta^2\gamma(t)^2}\; .
\label{doppler}
\end{equation}
The total AG is the sum over CBs (or large
individual GRB pulses) of the flux of Eq.~(\ref{Fnuobser}).

For an ISM of constant baryon density $\rm n_p$,
the deceleration of a CB  results in a Lorentz factor, $\rm\gamma(t)$,
that is given by (Dado et al. 2002a):
\begin{eqnarray}
\rm \gamma&=&\rm\gamma(\gamma_0,\theta,x_\infty;t)
=\rm {1\over B} \,\left[\theta^2+C\,\theta^4+{1\over C}\right]\;
,\nonumber\\
\rm C&\equiv&\rm
\left[{2\over B^2+2\,\theta^6+B\,\sqrt{B^2+4\,\theta^6}}\right]^{1/3}\; ,
\nonumber\\
\rm B&\equiv&\rm
{1\over \gamma_0^3}+{3\,\theta^2\over\gamma_0}+
{6\,c\, t\over  (1+z)\, x_\infty}\; ,
\label{cubic}
\end{eqnarray}
where $\gamma_0=\gamma(0)$, and
$\rm x_\infty=N_{_{CB}}/(\pi\, R_{max}^2\, n_p)$
characterizes the CB's slow-down in terms of $\rm N_{_{CB}}$, its
baryon number and $\rm R_{max}$, its radius (it takes a distance
$\rm x_\infty/\gamma_0$, typically of ${\cal{O}}(1)$ kpc, for the
CB to slow down to half its original Lorentz factor).

The extinction, $\rm A(\nu,t)$ in Eq.~(\ref{Fnuobser}), can be
estimated from the difference between the observed spectral index
{\it at very early time when the CBs are near the SN} and that
expected in the absence of extinction ($\rm F_{obs}\propto \nu^{-0.5}$,
for $\rm \nu\ll \nu_b$). The CB model predicts ---and the data
confirm--- the gradual evolution of the effective optical spectral
index towards the constant value $\approx -1.1$ observed in all
``late'' AGs (Dado et al. 2002a). The ``late'' index is independent
of the attenuation in the host galaxy, since at $\rm t>1$ (observer's)
days after the explosion, the CBs are typically moving in the
optically-transparent halo of the host galaxy.

The comparison of the predictions of Eq.~(\ref{Fnuobser}) with the
observations of optical, X-ray and radio light-curves and spectra
is discussed in Dado et al. (2002a,b).  The results ---for
{\it all} GRBs of known redshift--- are very simple, satisfactory
and parameter-thrifty. The CB-model results concerning X-ray lines
in GRB AGs  are also exceptionally predictive, simple and encouraging
(Dado et al. 2002e).

\section{GRB 021004 in the CB model}

Five properties of AGs in the CB model are particularly relevant to
GRB 021004:

\noindent
a. For most GRB optical AGs, good fits are obtained by approximating
the ISM-density by a constant (Dado et al. 2000a,b,c,d).  One
exception is the optical AG of GRBs ``caught'' at very early times,
when the CBs are still moving in a wind-generated density profile
$\rm \propto 1/r^2$. In the CB model the AGs' flux is proportional
to the instantaneous swept-in electron density, implying $\rm
F_{obs}\propto 1/t^2$ at very early times, as observed in GRB 990123
and GRB 991216.

\noindent
b. The achromatic ``bumps'' in the AGs of GRB 970508 and GRB 000301c are
explained (Dado et al. 2002a,b) by inhomogeneities in the density along
the CBs' trajectory, better than by gravitational lensing (Garnavich, et
al. 2000). Density changes are expected within star formation regions and
upon exit from the superbubbles where most of the SNe take place (Dado et
al. 2000a).

\noindent
c. The data on past AGs were course enough or started late enough for
the  contributions of different CBs (which can often be resolved as
individual pulses in the GRB phase) to coalesce into an AG contribution
describable as a single CB or a collection of similar ones.  But
individual CBs have different properties, and may even be emitted
at somewhat different angles, due to precession of the accretion
disk relative to the rotation axis of the compact object (Fargion
and Salis 1995), as observed in the microquasar  SS 433 (Margon
1984).

\noindent
d. The spectral shape of the AGs and their time-evolution,
in particular  their
steepening at the time-varying frequency $\rm\nu_b(t)$ of Eq.(\ref{nubend})
---towards $\nu^{-1.1}$ at late time--- is
very well supported by the data (Dado et al. 2002b).

\noindent
e. The excess polarization of AGs above that induced by the
ISM in our Galaxy may be largely due to
the host galaxy's ISM. In that case, it should be correlated with the
extinction in the host and decline with time as the CBs  move into the
halo.

The $\gamma$-ray light curve of GRB 021004 shows two prominent pulses
separated by $\sim\! 30$ seconds (Shirasaki et al. 2002 and
http://space.mit.edu/HETE/Bursts/GRB021004/). In the CB model, these
correspond to two dominant CBs.  The good quality of the optical data for
GRB 021004 ---and its double-humped $\gamma$ burst--- make irresistible
the temptation to fit its broad-band AG light curves with {\bf two CBs},
emitted in the same direction $\theta$ relative to the
observer\footnote{We have also made fits with two
different values of the emission angles $\theta$. They do not improve
$\chi^2$ significantly, and yield similar parameters: two different
emission angles are not really necessary.}, but with otherwise free
parameters (normalization, $\gamma_0$ and $\rm x_\infty$). We fix the
spectral index $\rm p$ in Eq.~(\ref{sync}) to the expected $\rm
p=2.2$, assume a density profile of the form $\rm n(r)=n_\infty\,
[(r_0/r)^2+1]$, and fit simultaneously all the reported
NIR, optical and radio data.

\begin{figure}[]
\vskip -1.5cm
\hskip 2truecm
\vspace*{- .4cm}
\plotone{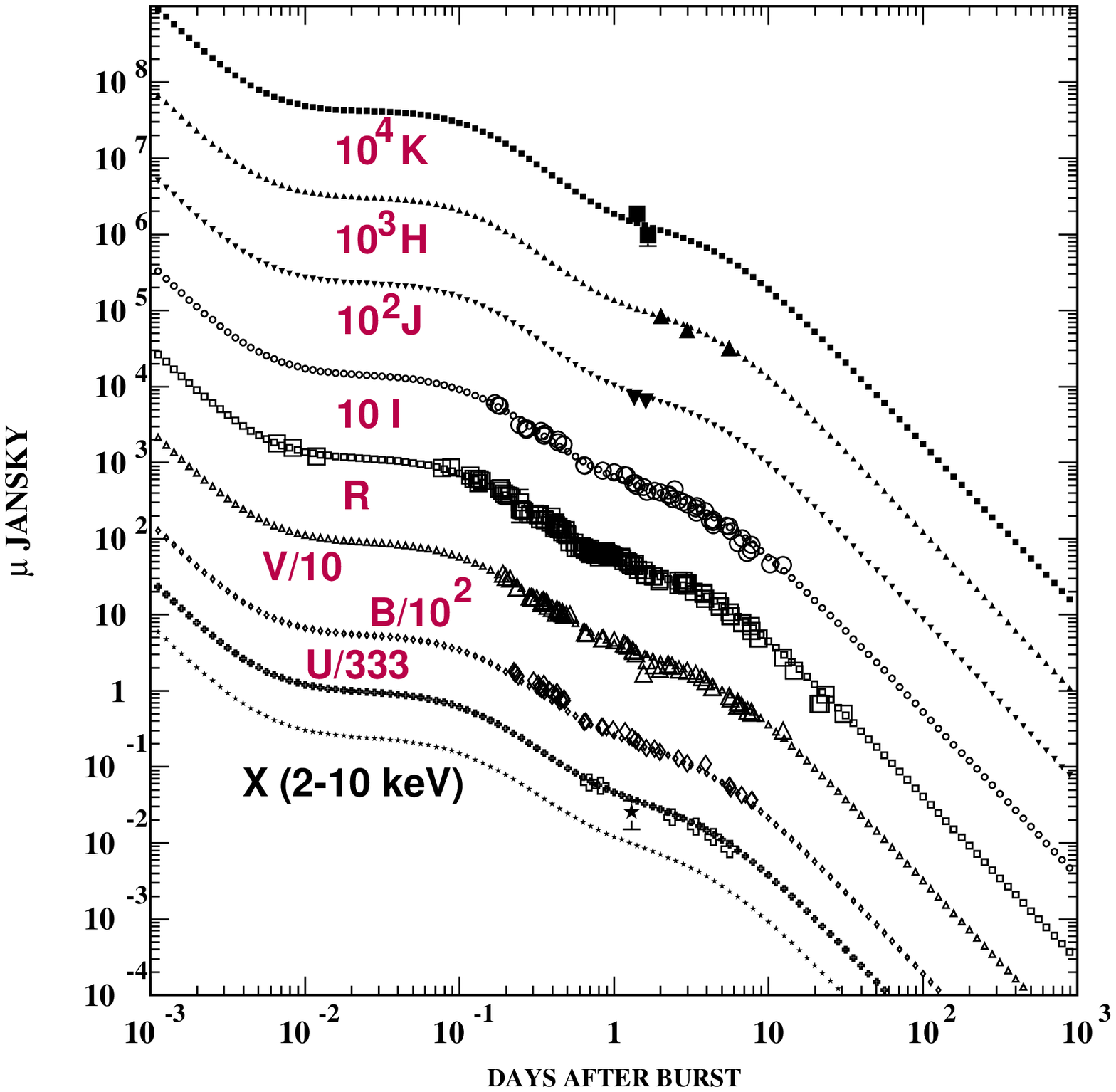}
\figcaption{Comparison between the NIR--optical
observations of the AG of GRB 021004 and the CB
model fit for two CBs with different
parameters, each contributing as in Eqs.~(\ref{sync}) to (\ref{cubic})
and corrected for Galactic, intergalactic and host extinction
as described in the text.
The ISM density was
assumed to be a constant plus an additional ``wind'' contribution
decreasing as $\rm 1/r^2$.
The figure shows (from top to bottom) $10^4$ times the K band,
$10^3$ times the H-band, $10^2$ times the J-band, 10 times the
I-band, the R-band, $10^{-1}$ of the V-band, $10^{-2}$ of the
B-band and $3\times 10^{-3}$ of the U-band.  The observational data
points are those reported to date, in
GCN notices (recalibrated with the observations of Henden et al. 2002),
and in Bersier et al. (2002), Holland et al. (2002) and Pandy et al.
(2002).
An estimated contribution of 0.66 $\mu$-Jansky of the host Galaxy,
(Holland et al. 2002) was subtracted from the late-time data
in the I, R and V bands.
The bottom line is the CB-model prediction for the {\it synchrotron
contribution} to the 2-10 keV X-ray AG (Datum is from Sako et al. 2002).
\label{fone}}
\end{figure}

\begin{figure}[]
\hskip 2truecm
\vspace*{0.8cm}
\vspace*{-.4cm}
\plotone{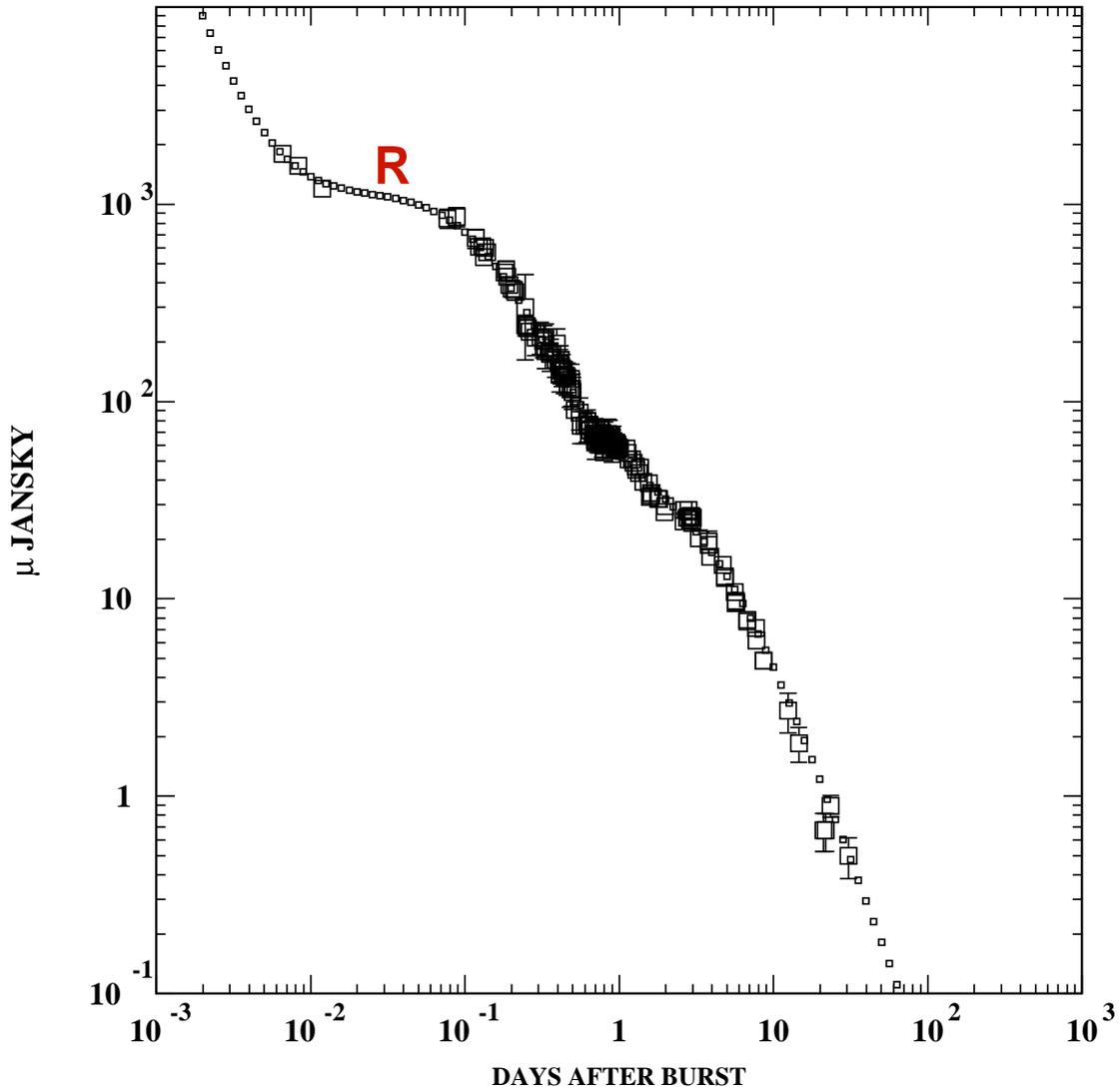}
\figcaption{Blow-up of the R-band results
of Fig. \ref{fone}.
The ISM density was
assumed to be a constant plus an additional ``wind'' contribution
decreasing as $\rm 1/r^2$. The wind contribution is
only significant at $\rm t\!<\!0.01$ days, after which the
CBs are more than 1 pc away from the progenitor.
The ``wind'' is also seen in other AGs observed early enough
(Dado et al. 2002a).
\label{ftwo}}
\end{figure}

\noindent
{\bf NIR-Optical AG:} In Fig. \ref{fone} we compare the results of
the CB-model's fit to the data for the NIR-optical lightcurves. We
have corrected for selective Galactic extinction in the direction
of GRB 021004 (Schlegel et al. 1998): $\rm E(B-V)=0.06$  (attenuation
magnitudes $\rm A_U=0.33,$ $\rm A_B=0.26,$ $\rm A_V=0.20,$ $\rm
A_R=0.16,$ $\rm A_I=0.12,$ $\rm A_J=0.08,$ $\rm A_K=0.04,$ and $\rm
A_H=0.01,$ in the various bands), and for absorption in the
intergalactic space and in the host galaxy in the B and U bands,
$\rm A_B=0.25$ and $\rm A_U=0.50$ mag,  estimated from the reported
spectra of GRB 021004 (Moller et al. 2002; Matheson et al.  2002;
Pandey et al. 2002; Bersier et al. 2002; Holland, et al. 2002;
Schaefer et al. 2002).  The contribution from the host galaxy to
the late AG, normalized to $0.66\pm 0.33$ $\mu$-Jansky in the R
band (Holland et al. 2002), was subtracted.  The results are very
satisfactory though, as befits a rough approximation to a no-doubt
very complex system, it is not perfect ($\chi^2=3.5 $ per d.o.f.
for 465 data points, if all reported statistical errors smaller
than 0.05 mag are fixed to 0.05 mag, to account for uncertainties
in calibration and in the conversion from magnitudes to Janskies).
The broad-band fitted parameters are, $\theta=1.47$ mrad,
$\gamma_0[1]=1403$, $\gamma_0[2]=1259$ (implying $\delta_0[1]=542$,
$\delta_0[2]=576$), $\rm x_\infty[1]=25$ kpc and $\rm x_\infty[2]=620$
kpc. To demonstrate the real quality of the fit, we have blown up
the R-band results in Fig. \ref{ftwo}. In the region between $\sim\!
0.5$ and $\sim\! 5$ days, the data ``wiggle'' by as much as 20\%
around the theoretical curve. It would be easy to correct for this
by assuming similar deviations of the ISM density relative to the
constant value adopted at large distances, clearly a moot exercise.

\noindent
{\bf Optical Polarization:} The observed polarization of the optical
AG of GRB 021004, which is consistent with that induced by the ISM
in our Galaxy (Covino et al. 2002a,b; Rol et al. 2002; Wang et al. 2002),
implies that the polarization induced by the host's ISM is rather small
and consequently the attenuation of the AG in the host must be small,
consistent with the CB model fit to the AG.

\noindent
{\bf X-ray AG:} The Chandra observations by Sako et al. (2002) in the 2-10
keV domain, from 0.867 to 1.874 days after burst, result in a
spectrum $\rm dn/dE\propto E^{-2.1\pm 0.1}$ and an AG decline $\rm dn/dt
\propto t^{-1.0\pm 0.2}$, both consistent with the optical observations
and with the CB-model's expectations: $\rm dn/dE\propto E^{-2.1}$ and,
except at earlier times, an achromatic light-curve.  In this GRB, as in
most others, the temporal and spectral behaviours are similar in the
optical and X-ray bands, as shown in Fig. \ref{fone}, but synchrotron
emission alone underestimates the X-ray fluence (in this case, the optical
to X-ray effective index is 0.2 units less steep than the separate
indices). The CB-model explanation is that in the X-ray domain not only
synchrotron radiation, but also Compton up-scattering and line-emission
 contribute to the flux (Dado et al. 2002a,b). There is
nothing exceptional in the X-ray data for GRB 021004.

\begin{figure}[]
\hskip 2truecm
\vspace*{0.8cm}
\plotone{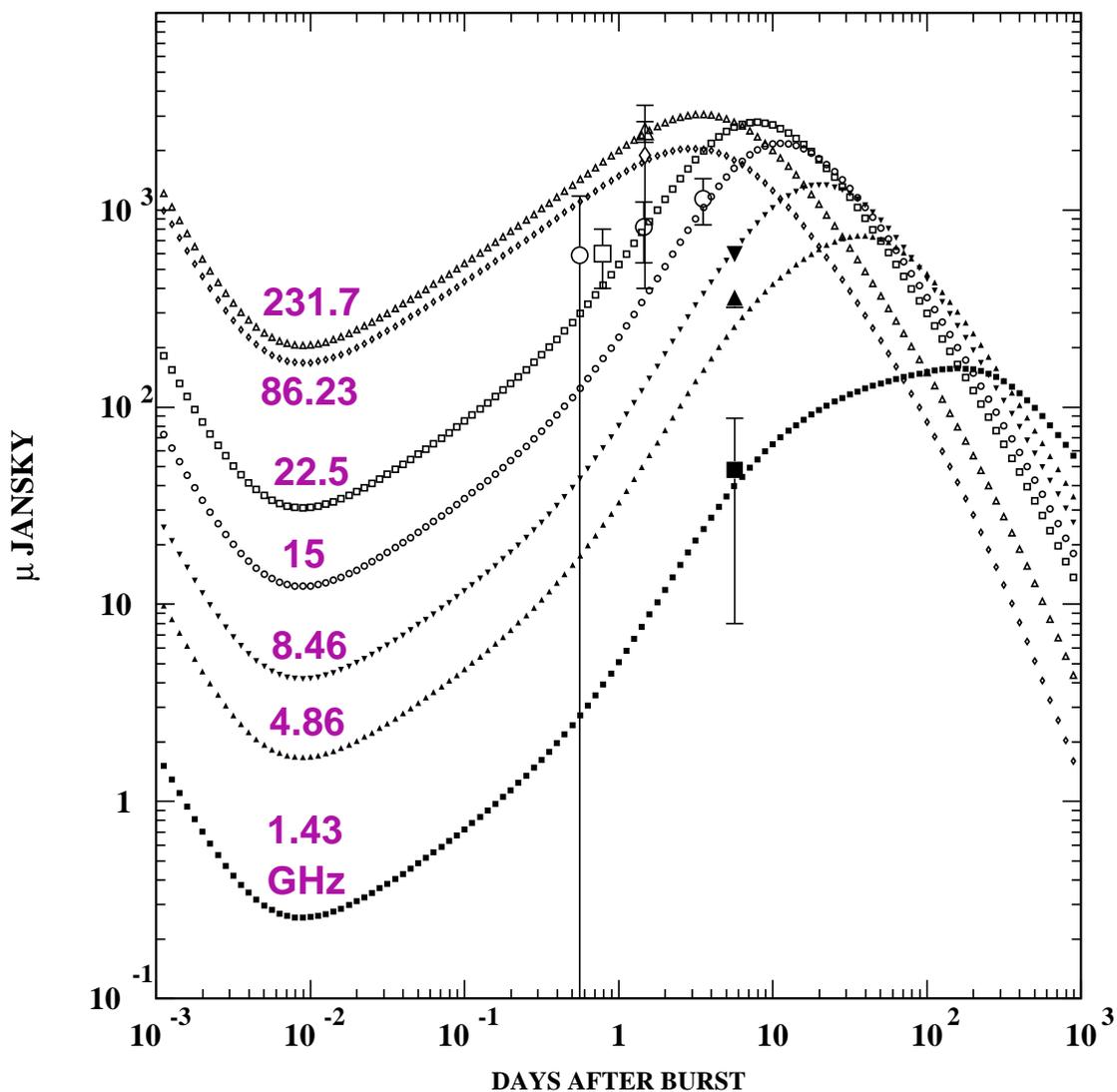}
\figcaption{Comparison between
the radio and
submillimeter observations of the AG of GRB 021004 and the CB-model
light-curves calculated from the broad band fit to its AG
with two CBs with different
parameters, each contributing as in Dado et al. (2002b).
The figure shows
(from bottom to top) the data and the predicted lightcurves at 1.43, 4.86,
8.46, 15, 22.5, 86.23 and 231.7 GHz. Data points are from Frail et al.
(2002), Berger et al.  (2002a,b);  Pooley et al. (2002a,b,c)
and Bremer et al. (2002). The symbols for the data are the same as those
for the corresponding theoretical curves.
\label{fthree}}
\end{figure}

\noindent
{\bf Radio and Submillimeter AG:} In Fig. \ref{fthree} we compare the
results of the CB-model's broad-band fit to the radio and submillimeter
data.  Unlike the X-ray and optical AGs, the radio AGs are sensitive to
self-absorption in the CBs, parametrized by a single absorption frequency
$\rm \nu_a$ (Dado et al.  2002b). The fit value for the CB that dominates
the ``late'' radio AG is $\rm \nu_a[2]=0.98$ GHz, similar to those of
other GRBs (Dado et al. 2002b).  The fit in Fig. \ref{fthree} is very
satisfactory, in particular in view of expected scintillations in the
radio data. The CB model is seen to correctly predict the early temporal
increase (to be followed by a later turn-over due to self absorption) and
the spectral behaviour ($\rm F_\nu\sim \nu^{-1.13\pm 0.15}$ between 1.43
GHz and 86 GHz, observed around 5 days after the burst) again akin to those
of other GRBs: 991208 (Galama et al. 2000), 000926 (Harrison et
al. 2001) and 000301c (Berger et al. 2000).  GRBs 021004 and 000301c also
have very similar redshifts and fluences, dim host galaxies, bumpy
light-curves and analogous ``late'' optical AGs (GRB 000301c was optically
detected after $\sim\! 2$ days by Fynbo et al.  2000).

\noindent
{\bf The GRB proper.} In the CB model (Dar and De R\'ujula 2000),
the rest-frame fluence of a CB viewed at a small $\theta$ is
amplified by a huge factor, $\delta_0^3$, due to the Doppler boost
and relativistic beaming of the radiation (Shaviv and Dar 1995).
In Dado et al.  (2002a) we deduced that the GRB photons of the GRBs
of known redshift correspond to a total energy release in the CBs'
rest system that is in a surprisingly narrow interval\footnote{GRBs
in the CB model are much better standard candles than in the standard
model (Frail et al.  2001).}, $10^{44\pm 0.3}$ erg; and
that the spread in the ``equivalent spherical energies'' $\rm
E_\gamma$ around their mean, $ \rm \bar E_\gamma\approx
4\times 10^{53}\, erg$, is mainly due to the spread of their Doppler
factors (deduced from the fits to their AGs) around their mean:
$\bar\delta_0\approx 1100$. For GRB 021004, $\delta_0\approx 550$ and
the CB-model expectation is  $\rm E_\gamma \approx \bar E_\gamma
(\delta_0/{\bar \delta_0})^3\approx 5\times10^{52}$ erg, in
agreement with the observed  $\rm \approx 4.8\times10^{52}$ erg
(we have corrected the value of Lamb
et al. (2002) to a redshift $\rm z=2.335$, from their
adopted $\rm z=1.6$).  The GRB of GRB 021004 is also entirely
normal.

\noindent
{\bf Spectral Variability:}
The energy dependence of the injection bend implies that the optical
spectra, $\rm F_{obs}[\nu,t]$, typically steepen in the first few
days, as the bend frequency in the observer's frame
$\rm\nu_b^{obs}(t)=\nu_b(t)\,\delta(t)/(1+z)$ ``crosses'' the
optical band   (Dado et al. 2002b).  Let
$\rm\lambda_b^{obs}(t)=c/\nu_b^{obs}(t)$. Given the parameters of
our fit to the optical AG, we can predict, by use of Eqs.(\ref{sync}),
the spectral energy density $\rm F_\lambda$ as a function of the
ISM density $\rm n_p$.  The two CBs of GRB 021004 have similar
$\gamma_0$ and $\delta_0$ but different $\rm x_\infty$:  their time
evolution and injection bends are different.  The AG is dominated
before and after $\rm t\sim 0.5$ days by one or the other CB, whose
properties determine the ``early'' or ``late'' spectral evolution.
The predictions for the spectral ratio at the ``late'' times ($\rm t_1=
0.756\,d,\, t_2=2.786\,d$) measured by Matheson et al. (2002),
are shown in Fig. \ref{ffour}.  The
upper panel is for $\rm n_\infty=0.0018\,  cm^{-3}$, the density
predicted from  $\rm x_\infty[2]$ using the CB-model reference values
($\rm N_{_{CB}}=6\times 10^{50}$,
$\rm R_{max}=2\times 10^{14}\, cm$; Dado
et al. 2000a,b). For this $\rm n_\infty$,
$\rm\lambda_b^{obs}(t_1)=3140$ \AA \ and
$\rm\lambda_b^{obs}(t_2)=8940$  \AA.

\noindent The interpolation between the limiting behaviours at $\rm
\nu\!\ll\! \nu_b$ and $\rm \nu\!\gg\! \nu_b$ could be more abrupt than in
Eq.~(\ref{sync}). In the lower panel of Fig. \ref{ffour} we illustrate the
result for a sharp transition, for which, with $\rm n_\infty=0.0048$
cm$^{-3}$,
$\rm\lambda_b^{obs}(t_1)=1932$ \AA \ (below the measured range) and
$\rm\lambda_b^{obs}(t_2)=5500$ \AA, at which point $\rm F_\lambda$ changes
from $\rm\sim [\lambda/\lambda_b]^{-1.5}$ to $\rm\sim
[\lambda/\lambda_b]^{-0.9}$, so that the ratio shown in the figure is $\rm
[\lambda/\lambda_b(t_2)]^{-0.6}$ below 5500 \AA, and unity above.

\section{Conclusions}

We have shown that the CB-model's interpretation of {\bf all the
data} concerning GRB 021004 is quite simple and very successful.
The only novelty is that, relative to other GRB AGs,  the
optical data start very early and are so copious
that a single CB would not fare well in explaining the
light-curve shapes. Two large pulses are observed
in the GRB's $\gamma$-ray light curve, enticing one to fit the rest of
the data
with two CB contributions. The results are excellent
and do not require new ad-hoc assumptions about the circumburst
ISM density: GRB 021004 fits well and naturally along all other
GRBs of known redshift.

\begin{figure}[]
\hskip 2truecm
\vspace*{0.8cm}
\plotone{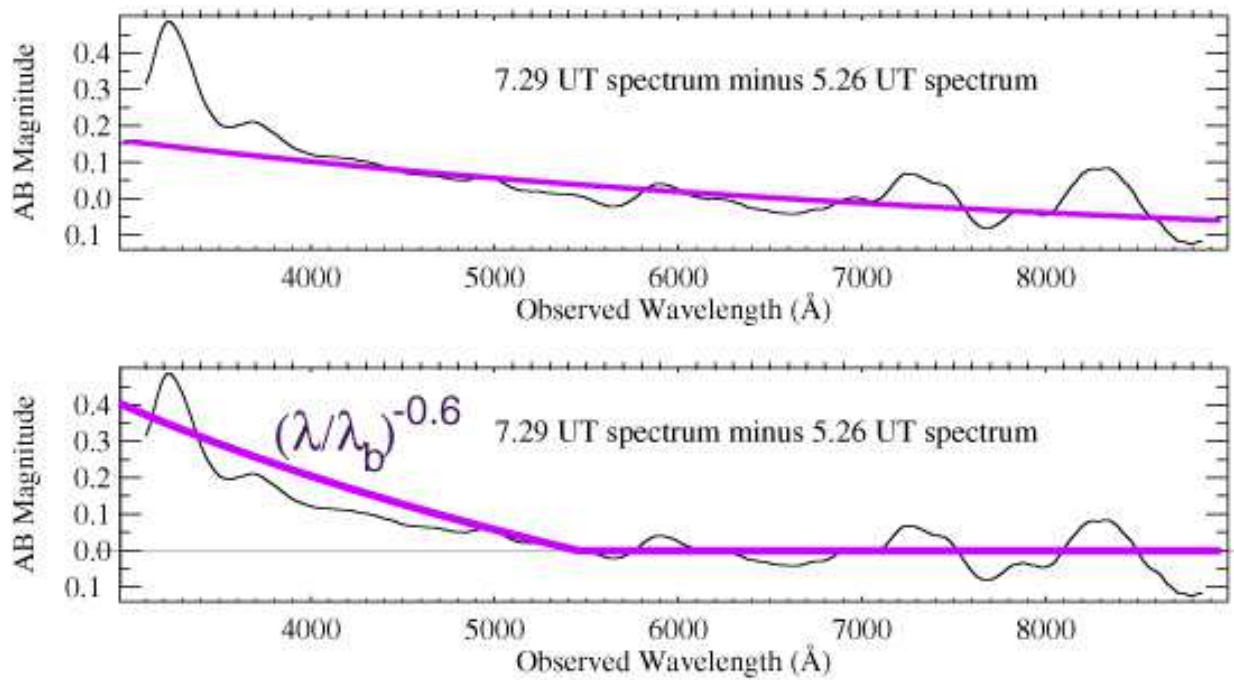}
\figcaption{Comparison between the ratio of the spectra ($\rm
F_\lambda $) of the optical afterglow of GRB 021004 at $\rm t_1=0.756$
days and $\rm t_2=2.786$ days after burst, normalized to $\sim\! 1$
at long wavelengths, as measured by Matheson et al. (2002). The
wiggles on the curve extracted from observations are artifacts of
the smoothing procedure, the error in the plotted magnitude difference
is estimated at 0.1 mag.  The smooth line in the upper panel is the
{\bf prediction}  from Eqs.~(\ref{sync},\ref{nubend}),
with the ISM density extracted from the observed $\rm x_\infty[2]$.
The smooth line in the lower panel
illustrates an abrupt injection bend at $\rm \lambda_b(t_2)=5500\, \AA$.
\label{ffour}}
\end{figure}

Having described a relatively simple AG evolution with 8
parameters, we cannot claim that they are all uncorrelated and
well determined, although their use results in the impressive
 prediction for the spectral evolution reported in Fig. \ref{ffour}.
 We regard our results as an existence proof of a
very simple CB-model understanding of the bumpy AG of GRB 021004.

It is, as usual, instructive to compare the CB-model's results to
those of the standard ``fireball'' scenarios. Two of the most
striking differences are their degree of simplicity and predictive
power; the explanation of the alleged X-ray lines in GRB afterglows
is an extreme case\footnote{In the CB model the existence and
time-dependent energies of the lines are predicted, requiring no
custom-made metallic envelopes, nor any other ad-hoc assumption
(Dado~ et al. 2002e).}. The GRB under discussion is also in an
extreme class: its standard explanations (Lazzati et al. 2002;
Nakar et al. 2002; Heyl and Perna 2002) require  model-dependent
inversions of the observed light curves to obtain the input functions
(the ISM-density or energy-supply profiles and the extinction in
the host) needed to ``predict'' the output functions (the light
curves at different frequencies). The various quoted authors extract
very different input functions, thereby illustrating the limited
``posdictive'' power of fireballs.  It would be difficult, on these
grounds,  to falsify any model with that much freedom.

For a non-constant ISM density, short-time spectral variations and
AG features are expected in the CB model due to the local instantaneous
dependence of $\rm F_\nu$ and $\rm\nu_b$ on the density,
Eqs.~(\ref{sync},\ref{nubend}). These are
not expected in the standard models because signals emitted at a
fixed $\rm t$ from different portions of the fireball or firecone
are observed at different times and with different Doppler boosts,
erasing all effects of width $\rm \Delta t\!\ll\! t$. Moreover, the
standard AG fluence depends on the integrated density (up to time
$\rm t$): an abrupt rise of fluence such as that observed in GRB
970508 requires a different explanation (the AG is ``re-energized'',
as in Piro et al. 1998). A similarly abrupt dip would be unexplainable.

The CB model is very modest in the adjectives that refer to GRBs.
None of them is exceptional, not even the very energetic GRB 990123,
nor 970508 with its peculiar AG shape, nor the extraordinarily close-by
980425. They are all associated with supernovae, seen when they are visible,
not seen when they are not (Dado et al. 2002a).
The explosions that generate GRBs are not ``the biggest
after the Big Bang''. The mechanism that begets GRBs is common:
it takes place in quasars and microquasars as well. The model works
very well and is very predictive, thus falsifiable.

{\bf Acknowledgment:} This research was supported in part by the
Helen Asher Space Research Fund and by the VPR fund for research
at the Technion. Useful remarks by Stephen Holland are
gratefully acknowledged.

\end{document}